\documentstyle[12pt]{article}
\begin{document}
\begin{center}
{\large \bf
$Q^2$ dependence of chiral-odd twist-3 distribution
$e(x,Q^2)$\\}

\vspace{5mm}
Y. Koike and N. Nishiyama
\\
\vspace{5mm}
{\small\it
Graduate School of Science and Technology, 
Niigata University, Niigata 950-21, Japan
\\ }
\end{center}

\begin{center}
ABSTRACT

\vspace{5mm}
\begin{minipage}{130 mm}
\small
We discuss the $Q^2$ dependence of the chiral-odd twist-3 
distribution $e(x,Q^2)$.  The anomalous dimension matrix for the 
corresponding twist-3 operators is calculated in the one-loop level.  
This study completes the calculation of the anomalous dimension 
matrices for all the twist-3 distributions together with the 
known results for the other twist-3 distributions $g_2(x,Q^2)$ 
and $h_L(x,Q^2)$.  We also have confirmed that in the large 
$N_c$ limit the $Q^2$-evolution of $e(x,Q^2)$ is wholely governed 
by the lowest eigenvalue of the anomalous dimension matrix which 
takes a very simple analytic form as in the case of $g_2$ and $h_L$.
\end{minipage}
\end{center}

\noindent
{\bf 1. Introduction}

\vspace{0.2cm}

The nucleon has three twist-3 
distributions $g_2$, $h_L$ and $e$\,[1].  
$g_2$ is chiral-even and 
the other two are chiral-odd.   $e$ is spin-independent 
and the other two are spin-dependent. 
Compared to $e$, $g_2$ and $h_L$ have 
more chance to be measured experimentally 
since they become leading contribution
to proper asymmetries in the polarized 
deep inelastic scattering (DIS) and Drell-Yan process,
respectively.  
Their $Q^2$ evolution has been studied in [2,3] 
for $g_2$ and in [4,5] for $h_L$.

Similarly to the distribution functions, there are three twist-3 
fragmentation functions which describes hadronization 
processes of partons 
in semi-inclusive processes\,[6], $\hat{g}_2$, 
$\hat{h}_L$
and $\hat{e}$. (Their naming is parallel to the corresponding 
distribution functions.)  In the inclusive  
pion production in the transversely polarized DIS,  
the chiral-odd fragmentation function $\hat{e}$ of the pion
appears as a leading contribution together with $h_1$ (twist-2)
of the nucleon. 
Although the $Q^2$ 
evolution of the twist-2 fragmentation functions is 
known to be obtained from
that of 
the corresponding distributions in the one-loop level (Gribov-Lipatov 
reciprocity\,[7]), no such 
relation is known for the higher twist 
fragmentation functions.

In our recent work\,[8], 
we have investigated the $Q^2$ evolution of $e(x,Q^2)$.  
Theoretically, this completed the calculation of the 
anomalous dimension 
matrices of all the 
twist-3 distributions.  Phenomenologically, we expect 
it will shed light on the $Q^2$ evolution of $\hat{e}(z,Q^2)$, 
anticipating the day when their relation is clarified.
This talk is a brief summary of [8].

\vspace{0.4cm}

\noindent
{\bf 2. $Q^2$ evolution of $e(x,Q^2)$}

\vspace{0.2cm}

The chiral-odd twist-3 distribution function 
$e(x,Q^2)$ 
is defined by the relation\,[1]
\begin{equation}
\int { d\lambda  \over 2 \pi } e^{i\lambda x} \langle P |
\bar{\psi}(0)  
\psi(\lambda n)|_Q |P \rangle 
= 2 Me(x,Q^2),
\end{equation}
where $|P\rangle$ is the nucleon (mass $M$)
state with momentum $P$. 
Two light-like vectors $p$ and $n$ defined by the relation,
$P=p+{M^2\over 2}n$, $p^2=n^2=0$, $p\cdot n=1$,
specify the Lorentz frame of the system.
Gauge-link operators are implicit in (1).
Taking the moments of (1) with respect to $x$, one 
can express the moments of $e(x,Q^2)$ in terms of
the nucleon matrix elements of the twist-3 operators
$V^{\mu_1\cdots\mu_n}$ (at $m_q=0$):
\begin{eqnarray}
{\cal M}_n\left[e(Q^2)\right] &=& e_n(Q^2),\\
\langle PS | V^{\mu_1 \mu_2 \cdots \mu_n} | PS \rangle &=& 2 e_n M 
(P^{\mu_1}P^{\mu_2}\cdots P^{\mu_n} -{\rm traces}),\\
V^{\mu_1 \cdot\cdot\cdot \mu_n} 
&=& \sum^{n}_{l=2} U_{l}^{\mu_1 \cdot\cdot\cdot \mu_n}
+ E^{\mu_1 \cdot\cdot\cdot \mu_n},\\
U_l^{\mu_1 \cdot\cdot\cdot \mu_n} 
&=& {1 \over 2}
{\cal S}_n \bar{\psi}\sigma^{\alpha\mu_1} 
 iD^{\mu_2}\cdots gG^{\mu_l}_{\ \, \alpha}
\cdots iD^{\mu_n} \psi - {\rm traces},\\
E^{\mu_1
\cdot\cdot\cdot \mu_n} 
&=& {1 \over 2} {\cal S}_{n}
\left[ \bar{\psi}i\rlap/{\mkern-1mu D} 
\gamma^{\mu_{1}}
iD^{\mu_{2}} \cdots iD^{\mu_{n}} \psi\right. \nonumber\\
& &\left. \qquad +
\bar{\psi} \gamma^{\mu_{1}}
iD^{\mu_{2}} \cdots iD^{\mu_{n}}i\rlap/{\mkern-1mu D}
\psi \right]
- {\rm traces},
\end{eqnarray}
where 
$ {\cal M}_n\left[ e(Q^2)\right]\equiv 
\int_{-1}^1\,dx\,x^n e(x,Q^2)$ and
the covariant derivative 
$D_\mu = \partial_{\mu} -igA_\mu$ restores
the gauge invariance.

$U_l$ contains $G_{\mu\nu}$ explicitly,
which indicates that $e(x)$ represents the quark-gluon 
correlations in the
nucleon.  
$E$ is the EOM (equation of motion) operator
which vanishes by use of the QCD equation-of-motion.
Although the physical matrix elements of EOM operators
vanish, one needs to take into account the mixing
with $E$ to carry out the renormalization of $U_l$,
as is discussed in [2,4] in the context of
the renormalization of $g_2$ and $h_L$.
From (4) one sees that $U_l$ appears in the form
of
\begin{eqnarray}
R_{n,l}^{\mu_1 \cdot\cdot\cdot \mu_n} 
= U_{n-l+2}^{\mu_1 \cdot\cdot\cdot \mu_n}
+ U_l^{\mu_1 \cdot\cdot\cdot \mu_n}, \ \ \ \left(
l=2,...,\left[{n \over 2}\right]+1 \right).
\end{eqnarray}
By this combination, $R_{n,l}$ has a definite charge 
conjugation $(-1)^n$.
Note the similarity between the present
\{$U_l$, $R_{n,l}$\} and \{$\theta_l$, $R_{n,l}$\} which appeared
in $h_L$\,[1,4].  
In fact the presence of 
$\gamma_5$ in $\theta_l$ is the mere difference from $U_l$.
$R_{n,l}$ of $h_L$ is defined as
$\theta_{n-l+2}-\theta_l$ and has a charge conjugation
$(-1)^{n+1}$ which is opposite to the above $R_{n,l}$ 
in (7).
We also recall that $e(x)$ does not mix with
the gluon-distribution owing to the chiral-odd nature.

For the renormalization of $e(x,Q^2)$, we choose
$R_{n,l}$ ($l=2,\ldots, \left[ \frac{n}{2} \right]+1$), $E$
as a basis of the operators.  
As in the case of $g_2$ and $h_L$, We eventually obtained 
the renormalization constants $Z_{ij}$ 
among  $R_{n,l}$, $N$ and $E$
in the following matrix form:
\begin{eqnarray}
\left(\matrix{R_{n,l}^B\cr
              E_n^B\cr}\right)=
\left(\matrix{Z_{lm}(\mu)&Z_{lE}(\mu)\cr
              0&Z_{EE}(\mu)\cr}\right)
\left(\matrix{R_{n,m}(\mu)\cr
              E_n(\mu)\cr}\right),
\ \ \ \left(l,m = 2,\cdot\cdot\cdot,\left[{n \over 2}
\right]+1\right).
\end{eqnarray}
This $Z_{ij}$ gives the anomalous dimension matrix for
\{$R_{n,l}$, $E$\}.

\vspace{0.4cm}

\noindent
{\bf 3. Large-$N_c$ limit}

\vspace{0.2cm}

In [3,5], it has been proved that
all the (nonsinglet) twist-3 distributions $g_2$, $h_L$ and $e$
obey a simple GLAP equation similarly
to the twist-2 distributions in the $N_c\to\infty$ limit.
In this limit, $Q^2$ evolution of these distributions
is completely determined
by the lowest eigenvalue of the anomalous dimension
matrix which has a simple analytic form.  For $e$, 
this can be also checked using the obtained result 
for the mixing matrix $Z_{ij}$.
At $N_c\to\infty$, i.e., $C_F\to N_c/2$, 
the lowest
eigenvalue of the anomalous dimension matrix for $R_{n,l}$ is
given by (ignoring the factor $g^2/8\pi^2$)
\begin{equation}
\gamma_n^e = 2N_c \left( S_n - \frac{1}{4} - \frac{1}{2(n+1)} \right).
\end{equation}
Furthermore the
$Q^2$ evolution of $e$ becomes
\begin{eqnarray}
{\cal M}_n\left[ e(Q^2)\right]=L^{\gamma_n^e/b_0}
{\cal M}_n\left[ e(\mu^2)\right].
\end{eqnarray}
As an example we compare the results for $n=4$.  The exact $Q^2$ 
evolution is 
\begin{eqnarray} 
{\cal M}_4\left[ e(Q^2)\right]
&=& \left(
0.983 a_{4,2}(\mu) + 0.520a_{4,3}(\mu) \right) 
\left({ \alpha(Q^2) \over
\alpha(\mu^2) }\right)^{9.59/b_0} \nonumber\\ &+& 
\left( 0.017 a_{4,2}(\mu) - 0.020a_{4,3}(\mu) \right) 
\left({ \alpha(Q^2) 
\over \alpha(\mu^2)
}\right)^{15.3/b_0},
\end{eqnarray} 
and the leading $N_c$ evolution is
\begin{eqnarray} 
{\cal M}_4\left[ e(Q^2)\right] = \left(
a_{4,2}(\mu) + \frac{1}{2}a_{4,3}(\mu) \right) 
\left({ \alpha(Q^2) \over
\alpha(\mu^2) }\right)^{10.4/b_0},
\end{eqnarray} 
where $a_{n,l}$ 
is
defined by
\begin{eqnarray}
\langle P | R_{n,l}^{\mu_1 \cdots \mu_n}(\mu^2) | P \rangle 
= 2a_{n,l}(\mu^2) M {\cal S}_n ( P^{\mu_1} 
\cdots P^{\mu_n} - {\rm traces}),
\end{eqnarray}
and $b_0={11\over 3}N_c -{2\over 3}N_f$.
One sees clearly that the coefficients in the 
second term of (11)
are small (i.e., $1/N_c^2$ suppressed) and the 
anomalous dimension in
(12) is close to the smaller one in (11).

\vspace{0.4cm}
\noindent
{\bf References}

\small
\vspace{0.2cm}

\noindent
[1] R.L. Jaffe and X. Ji, Nucl.\ Phys.\ {\bf B375}, 527 (1992).

\noindent
[2] J. Kodaira, K. Tanaka, T. Uematsu and Y. Yasui, 
hep-ph 9603377, and
references therein.

\noindent
[3] A. Ali, V.M. Braun and G. Hiller,  Phys. Lett. 
{\bf B266}, 117 (1991).

\noindent
[4] Y. Koike and K. Tanaka, Phys. Rev. 
{\bf D51}, 6125 (1995).

\noindent
[5] I.I. Balitsky, V.M. Braun, Y. Koike and K. Tanaka, 
hep-ph/9605439, Phys. Rev. Lett. in press.

\noindent
[6] R.L. Jaffe and X. Ji, Phys. Rev. Lett. {\bf 71}, 2457 (1993).

\noindent
[7] V.N. Gribov and L.N. Lipatov,  Sov. J. Nucl. Phys. 
{\bf 15}, 675 (1972).

\noindent
[8] Y. Koike and N. Nishiyama, hep-ph/9609207.


\end{document}